\documentclass[reprint,twocolumn,showpacs,superscriptaddress,preprintnumbers,10pt,amsmath,amssymb,
 aps,
pra]{revtex4-1} 

\usepackage{graphicx}
\usepackage{dcolumn}
\usepackage{bm}

\begin{document}

\title{Optical binding of  cylinder photonic molecules in the near-field of  partially coherent fluctuating Gaussian Schell model sources. \\ A coherent mode representation}

\author{Juan Miguel  Au\~{n}\'{o}n$^{1}$, F. J. Valvidia-Valero$^{1,2}$   and Manuel  Nieto-Vesperinas$^{*}$}

\affiliation{Instituto de Ciencia de Materiales de Madrid, Consejo Superior
de Investigaciones Cientificas, Campus de Cantoblanco, Madrid 28049,
Spain.
\\ 
$^{2}$Institut Carnot de Bourgogne, CNRS-UMR 5209, Universit\'{e} de Bourgogne, 21078 Dijon, France.\\
$^*$mnieto@icmm.csic.es\vspace*{0.1cm}}

\begin{abstract}
\begin{center}
Published in J. Opt.Soc. Am. A, {\bf 31}, 206-216, (2014).	
\end{center}We present a theory and computation method of radiation pressure from partially coherent light by establishing a coherent mode representation of the radiation forces. This is illustrated with the near field emitted from a  Gaussian Schell model source, mechanically acting on a single cylinder with  magnetodielectric behavior, or on a photonic molecule constituted by a pair of such cylinders. Thus after studying the force produced by a single particle, we address the effects of the spatial coherence on the bonding and anti-bonding states of two particles.  The coherence length manifests the critical limitation of the contribution of evanescent modes to the scattered fields, and hence to the nature and strength of the electromagnetic fores,  even when electric and/or magnetic partial wave resonances are excited.\end{abstract}

\pacs{350.4855, 260.2110, 030.6600, 030.1640.}

\maketitle 

\section{Introduction}
\label{introduction}
The subject of radiation forces from partially coherent light is receiving increasing attention \cite{wang2007effects, zha2009radiation,  korotkova2009twisted, gbur}. We recently put forward a systematic theory of photonic forces on small particles that characterized their magnitude by means of the diagonal elements of the cross-spectral density tensor \cite{aunon2012opticalforces}. A special aspect of this area has recently been recognized in connection with effects  from thermal sources such as those due to vacuum fluctuations, (Casimir, Van der Waals), and out of equilibrium forces \cite{Greffet, Antezza}, whose analogy with random near field forces from  partially coherent optical sources has been put forward \cite{aunon2012photonic, aunon2013photonic2}. In this connection, recent work deals with  interaction between two dipoles in presence of random wavefields \cite{Dogariu}.

On the one hand, the subject of optical forces is especially present in studies where light assisted mechanical interaction and nanomanipulation of  particles is of vital importance,  (see  Refs. \cite{Chaumet1, block2004review, povinelli, zemanek} and references therein). On the other hand, the statistical properties of radiation  introduces a new degree of freedom that plays a decisive role in optics \cite{mandel1995optical, james}: like e.g. in scattering processes \cite{Carney1,Lindberg2006,wolf2007introduction}, or speckle processing \cite{carminati2010subwavelength}, of relevance for the mechanical action of radiation beyond coherent light approaches \cite{Ashkin1970}.

In this work we  emphasize Schell model sources  whose spectral degree of coherence and radiant intensity distribution are both Gaussian \cite{mandel1995optical} (GSMS). They constitute an extraordinary instance of partially coherent  source that can be implemented in the laboratory without an excessive difficulty \cite{Gori2002synthesis}. Recently, the theory and consequences of the optical force generated by beams from this type of sources at far distances, or for general ABCD systems, where evanescent waves can be neglected, were reported  \cite{Friberg1988Imaging, korotkova2009twisted, aunon13equivalence}.  By contrast, in this work we address these forces in the near-field of the source.

For this purpose, we put forward a  theory of forces based on the concept of coherent mode representation (CMR) of partially coherent fields, due to E. Wolf, (see Refs. \cite{Wolf1982Coherentmode,Wolf1982newtheory}). This approach establishes that the cross spectral density of a system of any state of coherence  may be expressed as the sum of contributions from spatially completely coherent elementary sources, and so are its consequences for the electromagnetic force. We shall use this CMR of optical forces not only on single particles, but also for studying  radiation-induced forces between objects, usually referred to as optical binding \cite{Chaumet1,zemanek}. Specifically, we shall address the forces due to GSMS light, acting between two cylinders. We will exploit the morphology dependent resonances (MDR) of these objects to form different types of bonds between them. It will also be shown how the spatial coherence of the source affects the attraction or repulsion of these bodies. Although dielectric, the particles here addressed are also magnetic, namely they  respond   to the incident wave magnetic vector via induced magnetic dipoles and multipoles. They recently have provoked much interest because their potential as exotic scatterers capable of introducing configurations with artificial magnetism \cite{Peng2007_2, Vynck2009, GarciaEtxarri11Anistropic, magnelight}.

This paper is organized as follows: We briefly outline  in Section \ref{OF_GSMS} the theory of optical forces with partially coherent light emerging from a GSMS, with emphasis in the near-field. Then in Sections \ref{CMR} and  \ref{one_dimansional_GSMS} we develop the concept of stochastic forces from the point of view of the CMR.  Later, in the subsequent subsections, this method is analyzed and implemented through calculations of increasingly complex configurations. In Sections \ref{sec: Num_setup} and \ref{sec: 2particles} we apply this theory to the specific case of the mechanical action on a pair of magnetodielectric cylinders. An Appendix is added to show the  confirmation of our discussion.

\section{Optical Forces from Gaussian Shell Model Sources}
\label{OF_GSMS}
We shall consider Mie dipolar particles, namely those  whose scattering properties can be expressed in terms of the first electric and magnetic Mie coefficients \cite{nietoJOSA011,GarciaEtxarri11Anistropic}. Then the ensemble-averaged force experienced by the object is  decomposed into two contributions : a gradient (or conservative) force  ${\bf F}^{cons}$ and a non-conservative component ${\bf F}^{nc}$, which in terms of the electric vector  $\mathbf{E}(\textbf{r},\omega)$ at frequency $\omega$ reads \cite{nieto2004near,wong2006gradient,nieto2010optical,aunon2012opticalforces}:

\begin{eqnarray}
\left\langle F_{i}\left(\mathbf{r},\omega\right)\right\rangle  & = & \left\langle F_{i}^{cons}\left(\mathbf{r},\omega\right)\right\rangle + \left\langle F_{i}^{nc}\left(\mathbf{r},\omega\right) \right\rangle \nonumber \\
 & = & \frac{1}{4}\text{Re}\alpha\partial_{i} \left\langle  E_{j}^{*}\left(\mathbf{r},\omega\right)E_{j}\left(\mathbf{r},\omega\right)\right\rangle\nonumber\\
 &  & +\frac{1}{2}\text{Im}\alpha \left\langle E_{j}^{*}\left(\mathbf{r},\omega\right)\partial_{i}E_{j}\left(\mathbf{r},\omega\right) \right\rangle ,
\label{Fe}
\end{eqnarray}
where $(i,j)=(x,y,z)$, $\left\langle \cdot\right\rangle$ denotes ensemble averaged and $\alpha$ is the electric polarizability of the particle which characterizes the induced electric dipole: ${\bf p}(\textbf{r},\omega)=\alpha (\omega){\bf E}(\textbf{r},\omega)$ by the field emerging from the fluctuating source and impinging the particle. 

We next make use of the angular  spectrum  of plane waves $\textbf{e}\left(k\mathbf{s}_{\perp},\omega\right)$ \cite{nietolibro, mandel1995optical, nietoconj, nietolhm}:

\begin{equation}
\mathbf{E}\left(\mathbf{r},\omega\right)=\int_{-\infty}^{\infty}\mathbf{e}\left(k\mathbf{s}_{\perp},\omega\right)e^{ik\mathbf{s}\cdot\mathbf{r}}d^{2}\mathbf{s}_{\perp}.
\end{equation}
So that we can express the components of Eq. (\ref{Fe}) as \cite{aunon2012photonic}

\begin{eqnarray}
\left\langle F_{i}^{cons}\left(\mathbf{r},\omega\right)\right\rangle  & =-i\frac{k}{4}\text{Re}\alpha\iint_{-\infty}^{\infty}\textrm{Tr}\mathcal{A}_{jk}^{(e)}\left(k\mathbf{s}_{\perp},k\mathbf{s'}_{\perp}\omega\right)\nonumber \\
 & \times\left(s_{i}^{*}-s'_{i}\right)e^{-ik\left(\mathbf{s}^{*}-\mathbf{s'}\right)\cdot\mathbf{r}}d^{2}{\bf s}_{\perp}d^{2}{\bf s}'_{\perp},
\label{forcecons}
\end{eqnarray}

\begin{eqnarray}
\left\langle F_{i}^{nc}\left(\mathbf{r},\omega\right) \right\rangle  & =\frac{1}{2}\text{Im}\alpha\text{Im}\left\{ ik\iint_{-\infty}^{\infty}\textrm{Tr}\mathcal{A}_{jk}^{(e)}\left(k\mathbf{s}_{\perp},k\mathbf{s'}_{\perp}\omega\right)\right.\nonumber \\
 & \times\left.s'_{i}e^{-ik\left(\mathbf{s}^{*}-\mathbf{s'}\right)\cdot\mathbf{r}}d^{2}{\bf s}_{\perp}d^{2}{\bf s}'_{\perp}\right\},
 \label{forcenc}
\end{eqnarray}
where $k=\omega/c$,  $c$ being the velocity of light in vacuum. Also

\begin{equation}
\mathbf{e}\left(k\mathbf{s}_{\perp},\omega\right)=\frac{1}{(2\pi)^{2}}\int_{-\infty}^{-\infty}\mathbf{E}\left(\boldsymbol{\rho},\omega\right)e^{-ik\mathbf{s}_{\perp}\boldsymbol{\rho}}d^{2}\boldsymbol{\rho}.
\end{equation}
$\mathbf{E}\left(\boldsymbol{\rho},\omega\right)$ is the field at the exit plane $z=0$ of the source.

In these equations $\mathbf{s}=\left(\mathbf{s}_{\perp},s_{z}\right)$,  $\textbf{s}_{\perp}=(s_x,s_y)$,  and $s_{z}=\sqrt{1-s_{\perp}^{2}}$, when $|\textbf{s}|^{2}_{\perp}\leq1$;  or $s_{z}=i\sqrt{s_{\perp}^{2}-1}$ when $|\textbf{s}|^{2}_{\perp}>1$, which correspond to homogeneous and evanescent waves, respectively.  $\textrm{Tr}$ denotes the trace of the electric angular correlation tensor:  $\mathcal{A}_{jk}^{(e)}\left(k\mathbf{s}_{\perp},k\mathbf{s'}_{\perp},\omega\right)=\left\langle e_{j}^{*}(k\mathbf{s}_{\perp},\omega)e_{k}(k\mathbf{s'}_{\perp},\omega)\right\rangle $.

Now we address the specific case of a planar GSMS. This is characterized by a cross-spectral density tensor $W_{ij}^{(0)}\left(\mathbf{\boldsymbol{\rho}}_{1},\mathbf{\boldsymbol{\rho}}_{2},\omega\right)=\left\langle E_{i}^{*}\left(\boldsymbol{\rho}_{1}\right)E_{j}\left(\boldsymbol{\rho}_{2}\right)\right\rangle $ at the plane $z=0$  of the source  defined as \cite{mandel1995optical}
\begin{eqnarray}
&& W_{ij}^{(0)}\left(\mathbf{\boldsymbol{\rho}}_{1},\mathbf{\boldsymbol{\rho}}_{2},\omega\right)\nonumber\\
=&&\sqrt{S_{i}^{(0)}\left(\mathbf{\boldsymbol{\rho}}_{1},\omega\right)}\sqrt{S_{j}^{(0)}\left(\mathbf{\boldsymbol{\rho}}_{2},\omega\right)}
\mu_{ij}^{(0)}\left(\mathbf{\boldsymbol{\rho}}_{2}-\mathbf{\boldsymbol{\rho}}_{1},\omega\right),
\end{eqnarray}
where $S^{(0)}\left(\boldsymbol{\rho},\omega\right)=W^{(0)}\left(\boldsymbol{\rho},\boldsymbol{\rho},\omega\right)$ and  $\mu^{(0)}\left(\boldsymbol{\rho}_{1},\boldsymbol{\rho}_{2},\omega\right)$ are the spectral density and  the spectral degree of coherence of the source, respectively. In this model these quantities are both Gaussian, i.e.,
\begin{eqnarray}
S_{i}\left(\mathbf{\boldsymbol{\rho}},\omega\right)&=&A_{i}\text{exp}{[-\boldsymbol{\rho}^{2}/(2\sigma_{s,i}^{2})]} \\
\mu_{ij}\left(\mathbf{\boldsymbol{\rho}}_{2}-\mathbf{\boldsymbol{\rho}}_{1},\omega\right)&    =&B_{ij}\text{exp}[-(\boldsymbol{\rho}_{2}-\boldsymbol{\rho}_{1})^{2}/(2\sigma_{g,ij}^{2})].
\end{eqnarray}
$A_i$ is a constant, (equal to 1 in this work). The parameters $\sigma_{s,i}$ and $\sigma_{g,ij}$ are  the {\it spot size} and the {\it correlation} - or {\it spatial coherence} - {\it length},  respectively.

In this section, for simplicity, the electric field will be assumed to fluctuate in  the $Z-$direction,  so that  $B=1$. It is worth remarking that these fluctuations  along $OZ$, i.e. in  the direction of propagation, are negligible in the far-field; nevertheless  as we shall show, in the near-field they can be relevant and even larger that the rest of fluctuations.  In what follows we denote the parameters $\sigma_{i,s}$ and $\sigma_{ij,g}$ without the Cartesian subindex, understanding  that they refer to the $X$-component of the electric vector. 
\subsection{Near Field Forces}
\label{Near_Field_forces}

Let us address the optical forces of fields from  GSMSs on a small sphere, at distances from the source shorter than the wavelength. Whereas at larger distances, the trace of the angular correlation tensor can be approximated as $\text{Tr}{\cal A}_{ij} \simeq {\cal A}_{xx}$, in the near-field, where  the resolution of the system is  beyond the  diffraction limit: $\lambda/2$, the fluctuations on the $Z-$direction are as important as  the rest of them \cite{carminati1999near}. It is well-known that this conveys a non-straightforward 3D generalization in the definition of the degree of polarization $P(\textbf{r},\omega)$ \cite{setala2002near,Ellis2005degree,Ellis2005degree2,aunon2013degree}. 

Therefore, and in order to quantify the importance of these fluctuations we shall write  $e_z$   in terms of $s_x$, i.e., $e_{z}=-e_{x}s_{x}/s_{z}$, with the help of  the divergence law: $\mathbf{e}\left(k\mathbf{s}_{\perp}\right)\cdot\mathbf{s}=0$. Hence, $\text{Tr}{\cal A}_{ij}={\cal A}_{xx}+{\cal A}_{zz}$. The forces are calculated from Eqs. (1)-(8) on writing $(s_{x},s_{y})=s(\cos\theta,\sin\theta)$. The azimuthal  integrals are performed analytically, whereas the radial one is  numerically done for  $\sigma_g \gg \sigma_s$, this corresponds to a globally spatially coherent source. In this limit, the four integrals of the calculation can be expressed as a product of two integrals. We shall first consider a test particle with a radius $r_0=25nm$ and a relative permittivity $\varepsilon_p=2.25$. No  resonance effects appear in the chosen wavelengths. Then the dynamic electric polarizability that conserves energy \cite{nieto2010optical}, is $\alpha=4593+i17\;nm^3$. 

\begin{figure}[h!]
\begin{centering}
\includegraphics[width=\linewidth]{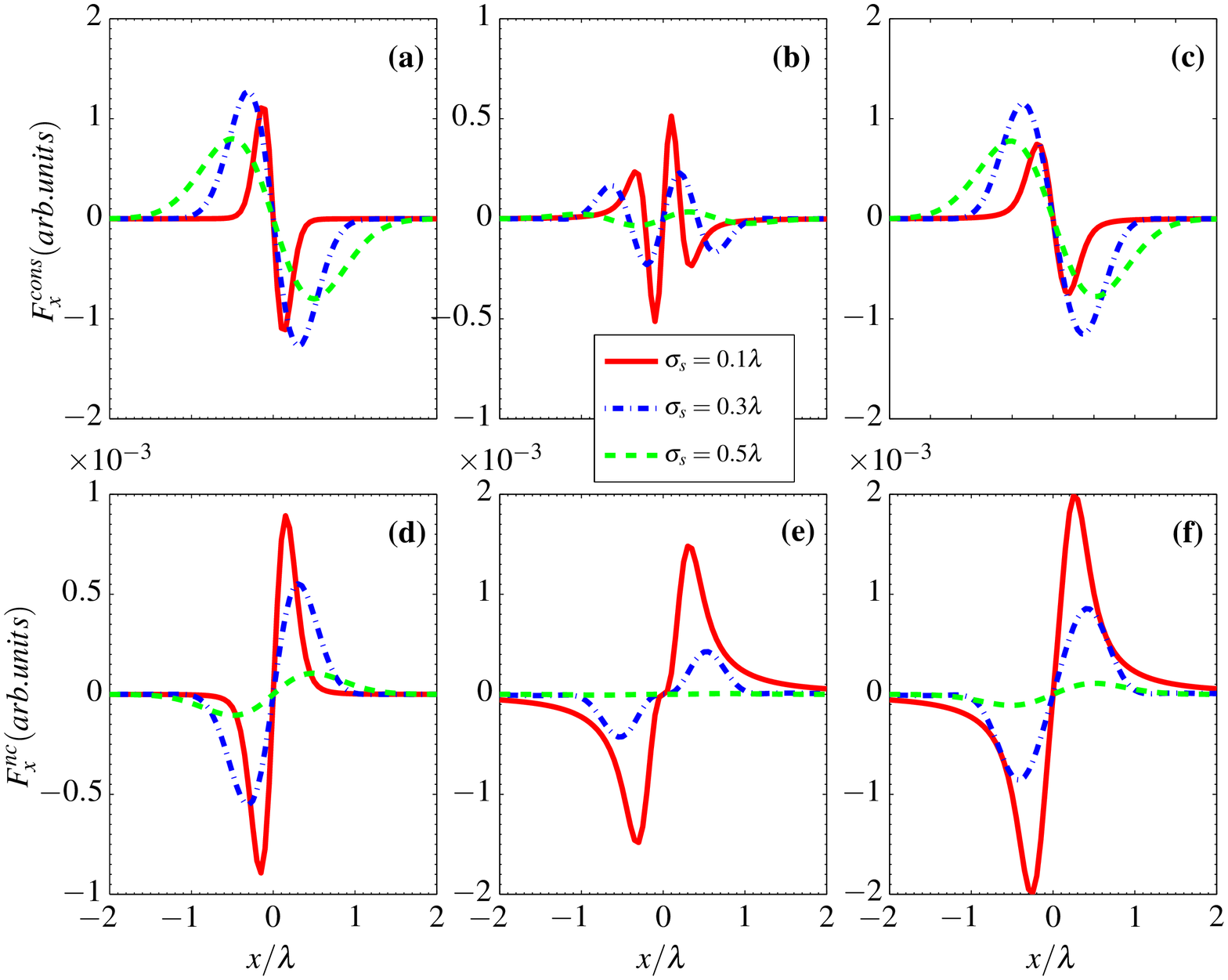}
\par\end{centering}
\caption{(Color online) . Mean forces. Conservative component $F_x^{cons}$, (first row), and non-conservative component  $F_x^{nc}$, (second row), of $F_x$ due to the contribution of $e_x$, (first column), and of $e_z$, (second column), versus the lateral displacement $x$  of the sphere,  (in wavelength units),  for different spot sizes $\sigma_s$. The third column displays the sum of the first and second columns. The distance of the particle to the source is $z=0.1\lambda$.
\label{fig_gradient}
}
\end{figure}

Fig. \ref{fig_gradient} shows the conservative force, (first row), and the non-conservative force, (second row), in the $X-$direction at a distance $z=0.1\lambda$. The contributions of the angular amplitudes $e_x$, (first column), and $e_z$, (second column), are separated. The third column is the sum of both forces.
\begin{figure}[h!]
\begin{centering}
\includegraphics[width=\linewidth]{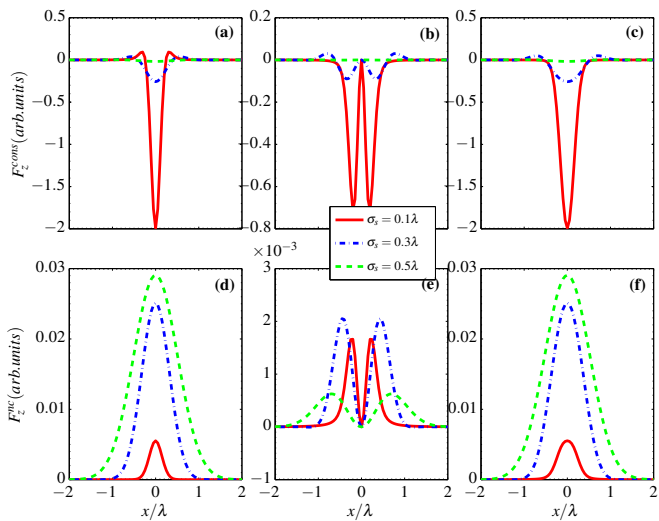}

\par\end{centering}
\caption{(Color online). The same as in Fig. \ref{fig_gradient} for $F_z$}
\label{fig_scatt}
\end{figure}
We see that all  contributions  of the components of $\mathbf{e}\left(k\mathbf{s}_{\perp},\omega\right)$ to $F^{cons}_x$ are of the same order, [compare the magnitude in Figs. \ref{fig_gradient}(a) and \ref{fig_gradient}(b) or \ref{fig_gradient}(d) and \ref{fig_gradient}(e)]; thus  the fluctuating $e_z$-s in the propagation direction  are not negligible like in the far-field, namely, $\text{Tr}{\cal A}_{ij} \not\approx {\cal A}_{xx}$. Fig. \ref{fig_scatt} shows the same as  Fig. \ref{fig_gradient}  but for the forces along the $Z-$axis. In this case we have the same effect as in the previous figure, although $F^{nc}_z$ is larger for the contribution of the $x-$fluctuations, [compare the magnitudes in Figs. \ref{fig_scatt} (d)-(e)]. By adding the conservative and non-conservative components of the force, i.e. Fig.\ref{fig_gradient}(c) and Fig.\ref{fig_gradient}(f), [and analogously for Fig. \ref{fig_scatt}], we see that the total force is only contributed by the gradient force, i.e., $F_x \simeq F_x^{cons}$ and $F_z \simeq F_z^{cons}$. This fact is due to the distance to the source being subwavelength.

These results also show that,  in general,  as  $\sigma_s$ increases, the magnitude of the forces decreases, contrary to far-field results of previous studies as well as to other configurations where the evanescent components do not play any role  \cite{wang2007effects,korotkova2009twisted,aunon13equivalence}. Notice that for $\sigma_s=0.1\lambda$ the beam condition:  $1/(2\sigma_{s})^{2}+1/\sigma_{g}^{2}\ll2\pi^{2}/\lambda^{2}$ (cf. \cite{mandel1995optical}) is not fulfilled, and it is precisely this value $\sigma_s$ that for which we obtain the largest magnitude of the force. Thus {\it the maximum force produced by a GSMS in the near-field corresponds to a minimum force in the far-field}. 

It should be pointed out that force calculations from a partially coherent source  are difficult without approximations, and are much more lengthy than those considered next.  Electromagnetic fields in complex structures are usually computed by  finite element methods (FEM) or by finite difference time domain procedures (FDTD). In Section \ref{CMR} and Section \ref{one_dimansional_GSMS} we develop a robust method in order to evaluate the cross spectral density tensor $W_{ij}\left(\mathbf{r}_1,\mathbf{r}_2,\omega \right) $, the degree of polarization $P(\textbf{r},\omega)$, and the optical forces $F(\textbf{r},\omega)$ in whatever set of particles. A test of this theoretical construction is shown in the Appendix, which confirms the results of Fig. \ref{fig_gradient} and Fig. \ref{fig_scatt}.

\section{Coherent mode representation}
\label{CMR}

The coherent mode representation (CMR) establishes that a stationary optical field of any state of coherence may be represented as a superposition of coherent modes \cite{Wolf1982newtheory,tervo2004theory}, i.e.,
\begin{eqnarray}
W_{ij}\left(\mathbf{r}_{1},\mathbf{r}_{2},\omega\right)&=&\left\langle E_{i}^{*}\left(\mathbf{r}_{1},\omega\right)E_{j}\left(\mathbf{r}_{2},\omega\right)\right\rangle \nonumber \\
&=&{\displaystyle \sum_{q}\lambda_{q}\left(\omega\right)\phi_{i,q}^{*}\left(\mathbf{r}_{1},\omega\right)}\phi_{j,q}^{}\left(\mathbf{r}_{2},\omega\right),
\label{W_sum}
\end{eqnarray}
where $\lambda_q(\omega)$ are the eigenvalues and $\phi_{q,i}$ denote the eigenfunctions which fulfill the equation \cite{Wolf1982newtheory}:
\begin{equation}
\int_{D}\phi_{i,q}\left(\mathbf{r}_{1},\omega\right)W_{ij}\left(\mathbf{r}_{1},\mathbf{r}_{2},\omega\right)d^{3}\textbf{r}_1=\lambda_{q}\left(\omega\right)\phi_{i,q}\left(\mathbf{r}_{2},\omega\right).
\label{W_int}
\end{equation}

Let us consider an statistical ensemble of electromagnetic fields $\lbrace\textbf{E}(\textbf{r},\omega)\rbrace$ where each realization can be expressed as a sum of individual eigenfunctions:
\begin{equation}
E_i(\textbf{r},\omega)=\displaystyle \sum_{q}a_q(\omega) \phi_{i,q}(\textbf{r},\omega),
\label{E_sum}
\end{equation}
 $a_q$ being a random coefficient. Substituting Eq. (\ref{E_sum}) into Eq. (\ref{W_sum}) we see that 
\begin{eqnarray}
\left\langle a_{q}^{*}\left(\omega\right)a_{q'}\left(\omega\right)\right\rangle &=&\lambda_{q}\left(\omega\right)\delta_{qq'}, \nonumber \\
a_{q}\left(\omega\right)&=&\lambda_{q}^{1/2}\left(\omega\right)e^{i\alpha_{q}},
\end{eqnarray}
where $\alpha_q$ is a real random variable uniformly distributed in the interval $0\leq \alpha_q<2\pi$.

\subsection{Coherent Mode representation of Optical forces}
\label{CMR_OF}

We  can now write the ensemble-averaged force as a  sum of coherent modes by  using its expression from the momentum conservation law in terms of the Maxwell stress tensor (MST) \cite{jackson1998classical,ChaumetOL,Cui2008}:

\begin{eqnarray}
& &\left\langle \mathbf{F}\left(\mathbf{r},\omega\right)\right\rangle  \nonumber\\  & =&\sum_{q}\iint_{\Sigma}\frac{\varepsilon}{2}\text{Re}\left\{ \left\langle \left(\mathbf{E}_{q}\cdot\mathbf{n}\right)\mathbf{E}_{q}^{*}\right\rangle \right\} -\frac{\varepsilon}{4}\left\langle \mathbf{E}_{q}^{*}\cdot\mathbf{E}_{q}\right\rangle \mathbf{n} \nonumber\\
&&+\frac{\mu}{2}\text{Re}\left\{ \left\langle \left(\mathbf{H}_{q}\cdot\mathbf{n}\right)\mathbf{H}_{q}^{*}\right\rangle \right\} -\frac{\mu}{4}\left\langle \mathbf{H}_{q}^{*}\cdot\mathbf{H}_{q}\right\rangle \mathbf{n} ds. \nonumber\\
\label{F_Tij}
\end{eqnarray}
$\Sigma$ is a  surface enclosing the object experiencing the force. ${\bf n}$ represents the outward unit normal. In our 2D calculations  $\Sigma$ will be a closed line.  ${\bf E}_q$, ${\bf H}_q$ and ${\bf E^\ast}_q$, ${\bf H^\ast}_q$ are the  $q$-modes and their complex conjugates. For brevity we have omitted the  space and frequency dependence of the fields. $\epsilon$ and $\mu$ are the permittivity and permeability of the surrounding medium embedding the particles, which in this work will be assumed to be vacuum.  The sum of the partial forces from each propagated eigenmode  renders the resulting force exerted on the particles by the total fields ${\bf E}$ and ${\bf H}$.  Notice that Eq. (13)  applies to any configuration, regardless of whether the source is spatially coherent ($q=0$) or partially coherent ($q>0$).

For dipolar particles the averaged  total force Eq. (1) can now be expressed in terms of the coherent  $q$-modes:
\begin{eqnarray}
\left\langle F_{i}\left(\mathbf{r},\omega\right)\right\rangle &=& \frac{1}{2}\sum_{q}\text{Re}\left\{ \alpha_{e}E_{j,q}\partial_{i}E_{j,q}^{*}\right\} \nonumber\\
&=&\frac{1}{2}\sum_{q}\lambda_{q}\text{Re}\left\{ \alpha_{e}\phi_{j,q}\partial_{i}\phi_{j,q}^{*}\right\}.
\end{eqnarray}

\section{Characterization of the field emitted by the GSMS}
\label{one_dimansional_GSMS}

Using the CMR, we shall follow the procedure put forward in  \cite{Lindberg2006} to characterize the fluctuating field from a GSMS . Then  the problem is 2D so that plane of work will be $XY$.  $y$ is the direction of propagation and the field  fluctuates along  $0Z$ (see Fig. \ref{fig: Numerical_setup}). The GSMS plane is  $y=0$, thus the cross-spectral density function will be:
\begin{equation}
W^{(0)}_{zz}\left(x_{1},x_{2},\omega\right)=Ae^{-\frac{x_{1}^{2}+x_{2}^{2}}{4\sigma_{s}^{2}}}e^{-\frac{\left(x_{1}-x_{2}\right)^{2}}{2\sigma_{g}^{2}}}.
\label{Wzz_onedim}
\end{equation}

For this case, the eigenfunctions and the eigenvalues have been determined previously \cite{mandel1995optical,Gori1980Colletwolf}:

\begin{equation}
\phi_{q}\left(x,\omega\right)=\left(\frac{2c}{\pi}\right)^{1/4}\frac{1}{\left(2^{q}q!\right)^{1/2}}H_{q}\left(x\sqrt{2c}\right)e^{-cx^{2}},
\label{eigenfunction}
\end{equation}

\begin{equation}
\lambda_{q}\left(\omega\right)=\left(\frac{\pi}{a+b+c}\right)^{1/2}\left(\frac{b}{a+b+c}\right)^{q},
\end{equation}
where $H_q(x)$ is the Hermite polynomial of order $q$,  and 
\begin{equation}
a=\frac{1}{4\sigma_{s}^{2}},\;\; b=\frac{1}{2\sigma_{g}},\;\; c=\left(a^{2}+2ab\right)^{1/2}.
\end{equation}

The angular amplitude $\Phi(ks_x)$ of the eigenfunction $\phi_q(x,\omega)$, is calculated  by inverse Fourier transform of  Eq. (\ref{eigenfunction}),  (see \cite{Lindberg2006}):

\begin{eqnarray}
\Phi(ks_{x})&=&\frac{1}{2\pi}\int_{-\infty}^{\infty}\phi\left(x,\omega\right)e^{-iks_{x}x}dx\nonumber\\
&=&\frac{(-i)^{q}}{2\pi}\left(\frac{2\pi}{c}\right)^{1/4}\frac{1}{(2^{q}q!)^{1/2}}e^{-\frac{k^{2}s_{x}^{2}}{4c}}H_{q}\left(\frac{ks_{x}}{\sqrt{2c}}\right). \nonumber\\
\label{eigenfunction_angular}
\end{eqnarray}

\section{Numerical setup}
\label{sec: Num_setup}

A pair of particles  is illuminated by the GSMS wavefield  whose mechanical action produces  optical binding effects  with characteristics of a photonic molecule. \cite{Boriskina2006,Boriskina2007,Boriskina2010}.  2D numerical calculations are done by means of a FEM (RF module of COMSOL 4.3a, \mbox{http://www.comsol.com}) and MATLAB. Aside from some depolarization effects, the main features of the physical process: light scattering, resonance excitation and binding, are analogous to those in 3D \cite{vandeHulst1981,Taflove2004,Ho1994}.

Without loss of generality, a {\it Si} cylinder with $\epsilon=10.24$ and radius $r_{0}=0.3\mu m$ \cite{Boriskina2007} has been considered, due to its rich Mie resonance spectra in both the visible and near IR \cite{ValdiviaValero2012_1}. This will allow us to analyze the effects of  spatial coherence on these  resonances and their consequences for the induced optical forces on this pair,  (see Section \ref{one_dimansional_GSMS}).  

\begin{figure}[htbp]
\begin{minipage}{\columnwidth}
\centering
\includegraphics[width=0.8\linewidth]{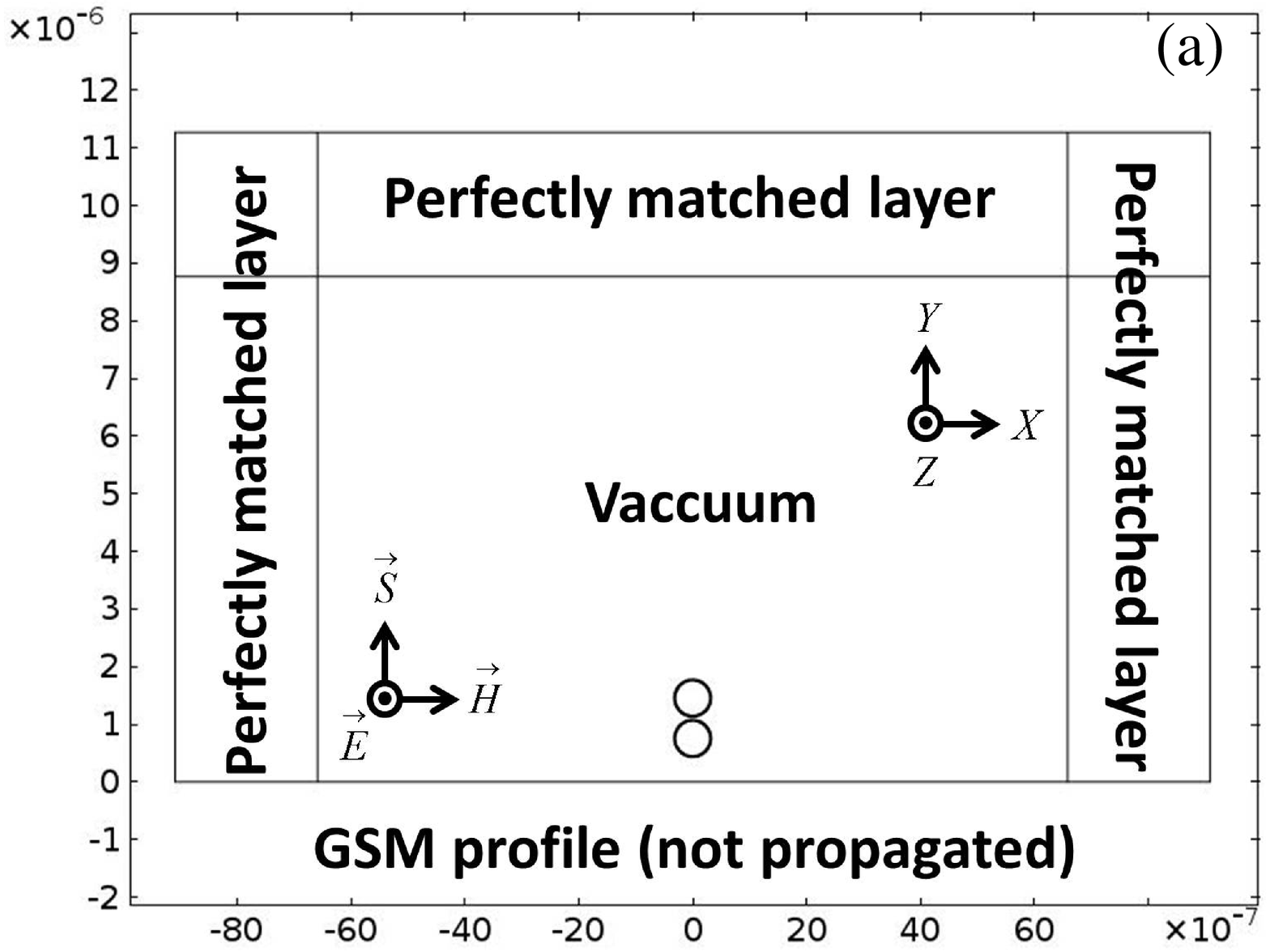}
\end{minipage}
\begin{minipage}{\columnwidth}
\centering
\includegraphics[width=0.8\linewidth]{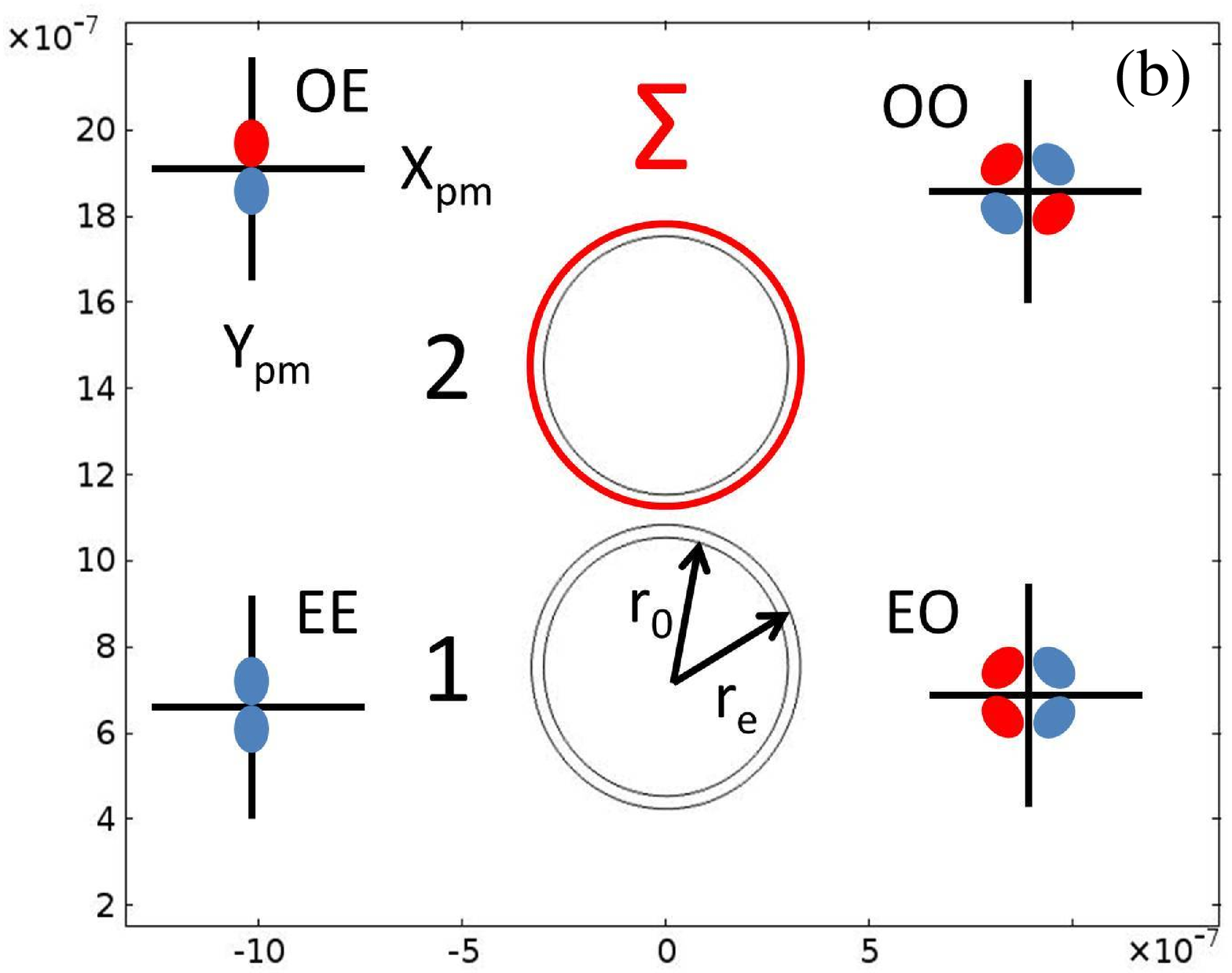}
\end{minipage}
\caption{(a) Illustration of the geometry for resonant  wavelength identification of both the single particle and the pair;  as well as for the computation of the optical forces. An incident S-polarized field with a GSMS profile, (amplitude $A = 1W/m^2$, width of its intensity $\sigma_s = 0.05\times 1500nm$, degree of coherence $\sigma_g=100\sigma_s,2\sigma_s,0.5\sigma_s$), impinges the {\it Si} cylinders of radius $r_{0}$ with excitation of  their  WGMs: $TE_{mn}$. (a) In order to simulate infinite space, three absorbent, or perfectly matched, layers (PML) are located at the upper and lateral boundaries of the calculation window; the lower boundary containing the incident wave profile  of the GSMS. (b) Detail of the geometrical cross sections of the particles conforming the {\lq\lq}photonic" molecule, where the light intensity $\left|\left\langle {\bf S}(\textbf{r})\right\rangle \right|$ is averaged to the surface of the cilinder of radius $r_0$, and the circumference $\Sigma$ of radius $r_{e}$ surrounding each  particle is employed  to calculate the electromagnetic forces (per axial  unit length), [cf. Eq. (13)], (see also \cite{ValdiviaValero2012_2}). Particles 1 and 2 stand for the lower/right, directly illuminated by the beam, and the upper/left ones, respectively.}
\label{fig: Numerical_setup}
\end{figure}

Following the scheme shown in Fig. \ref{fig: Numerical_setup}(a), an incident  wavefield with electric vector $E_{z}$ perpendicular to the $XY$-plane, is launched upwards, propagating along $OY$. The choice of S-polarization (TE), in contrast with P-polarization (TM), excites whispering-gallery modes (WGMs):  $TE_{mn}$, (where {\it m} and {\it n} indicate the angular and radial orders, respectively), which extend to the near field region surrounding the cylinders. This facilitates the electromagnetic interaction between these particles. The light directly illuminates  the right or lower particle, depending on whether the orientation of the molecular set is horizontal or oblique/vertical. Correspondingly, either the left or the upper cylinder is mostly excited by the resonance of its partner. This technique is performed  so that  the molecular states associated to anti-symmetric field patterns with respect to the transversal axis of the molecule, [see insets $OE$ and $OO$ in Fig. \ref{fig: Numerical_setup}(b)] are not destroyed, especially when the molecule is inclined  with respect to the propagation direction of the beam. Notice that if both particles were simultaneously illuminated by the beam, only those WGMs related to symmetric field patterns with respect to the molecule axis  would be excited, [see insets $EE$ and $EO$ in Fig. \ref{fig: Numerical_setup}(b)].

The separation between the particles is $d_0=100nm$, which makes subwavelength the molecule   dimensions, (compare the set size, $1.3\mu m$, to the range of wavelngths under study: $1.6\mu m - 8.0\mu m$). The center of the lower particle is $\approx 0.75\mu m$. We follow the nomenclature of  \cite{Boriskina2006,Boriskina2010} for the molecular states,  the classification being based on the ${\bf E(r)}$ field symmetry with respect to the main directions defined by the molecule geometry, i.e. its longitudinal ($Y_{pm}$) and transversal ($X_{pm}$) axes, [see Fig. \ref{fig: Numerical_setup}(b)]. As an example, we will examine the upper-left inset of Fig. \ref{fig: Numerical_setup}(b). In this case, the upper lobe is opposite to the lower one, thus it is said that $E$  is even, (E),  with respect to $OX$; however,  $E$  is odd. (O), with respect to $0Y$. Therefore the photonic state is even-odd (EO). If they mismatch one another, it would be $X_{pm}Y_{pm}/OO$, [cf.  the upper-right inset in Fig. \ref{fig: Numerical_setup}(b)]. These would be similar to a double bound in the molecule. In the case in which only one lobe of each particle interacts with the other (simple bound), the states will be $X_{pm}Y_{pm}/EE$ and $X_{pm}Y_{pm}/OE$, respectively.

In all cases the $E_z$ profile at frequency $\omega$ is that of a GSMS, described in Section \ref{one_dimansional_GSMS}. The field has an intensity $1W/m^2$ and $\sigma_s=0.05\times 1500nm$. The spatial coherence of the near field is gradually established as the ratio between the coherence length  and the width of the beam $\sigma_g/\sigma_s=100,2,0.5$ diminishes. The GSMS is placed in the lower boundary of the simulation window and is implemented as a discrete  sum of modes $q$,  [see Eq. (\ref{eigenfunction})]. As explained in Section \ref{one_dimansional_GSMS}, the lower the ratio of $\sigma_g/\sigma_s$, the higher the value of $q$, (cf.  Fig. 5.17 of \cite{mandel1995optical}). An iterative process is followed in order to simulate the propagation of each of these $q$-modes through the calculation window. Subsequently, they are summed up to get the propagated total fields ${\bf E(r)}$ and ${\bf H(r)}$.

The next results show the time-averaged energy flow $\left\langle {\bf S}(x,y)\right\rangle $, that shows light concentration in the probe cylinders. Because of their dielectric nature, we average $\left|\left\langle {\bf S}(x,y)\right\rangle \right|$ in a circle which coincides with the geometrical cross section of the probe cylinder of radius $r_{0}$, [see Fig. \ref{fig: Numerical_setup}(b)]. This stems from the fact that, if the particle is dielectric, the intensity of the light beam that couples to the particle WGM, is concentrated inside the cylinders, (see \cite{ValdiviaValero2010}), not outside them, (the latter occurs for plasmonic cilinders \cite{ValdiviaValero2011_3}). In all cases, these intensities are normalized to the maximum intensity of the incident Gaussian beam: $\left|\left\langle {\bf S}_{max}\right\rangle \right|= 1W/ m^2$.

The averaged force on the probe cylinders is calculated by employing the MST, Eq. (\ref{F_Tij}). The line of integration $\Sigma$  surrounds each particle as seen in Fig. \ref{fig: Numerical_setup}(b).  In our 2D geometry, $\Sigma$ is the circumference of radius $r_e$, (see Fig. \ref{fig: Numerical_setup}(b)).  $\epsilon=\mu=1$. Because of this 2D geometry, our results are expressed as force per axial length unit, in $N/m$.

The COMSOL calculation with  complex values of ${\bf E(r)}$ and ${\bf H(r)}$ as well as of the real physical fields: ${\bf E^R}({\bf r}, t) = \text{Re}[{\bf E(r)}\exp(-i\omega t)]$ and ${\bf H^R}({\bf r}, t) = \text{Re}[{\bf H(r)}\exp(-i\omega t)]$, is not straightforward. The details of the procedure have been given in \cite{ValdiviaValero2012_2}. The meshing used in the simulation has a maximum and a minimum element of $\lambda_{ref}/8$ and $2.7nm$, respectively. The reference wavelength being  $\lambda_{ref}=1620nm$. The maximum element growth rate, resolution of curvature, and resolution of narrow regions are   $1.3$, $0.3$, and $1$, respectively.

\section{A bi-particle  molecule illuminated by a GSMS beam. Effects of partial coherence in the  {\lq\lq}molecular" states}
\label{sec: 2particles}

\subsection{Localization of resonances of a single particle. Bi-particle set: Production of {\lq\lq}molecular" states}
\label{subsec: Intensity_2particles}

In order to identify the resonant states of a photonic molecule, the spectral  location  of the resonances of the single particle is required. For the sake of  accuracy  needed in the calculations, and  in order to deal with  not too complex bonds between the particles, our study limits the search of resonances in each individual particle to those of low angular order.  This suffices to illustrate the analysis in this work. 

\begin{figure}[htbp]
\begin{minipage}{\columnwidth}
\centering
\includegraphics[width=0.8\linewidth]{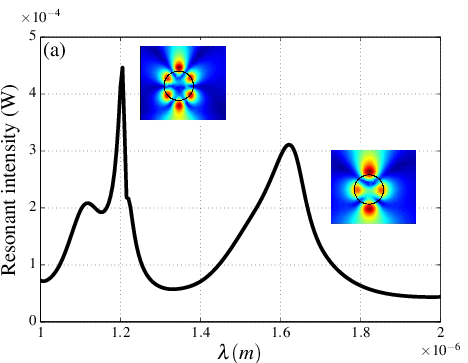}
\end{minipage}
\begin{minipage}{\columnwidth}
\centering
\includegraphics[width=0.8\linewidth]{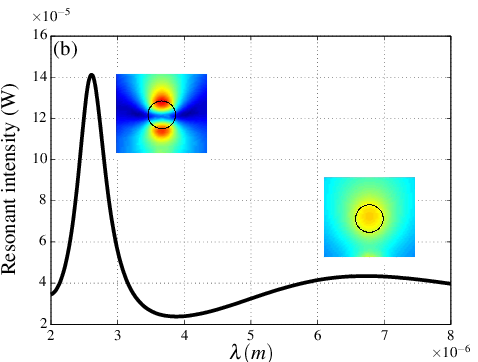}
\end{minipage}
\caption{Spatially coherent illumination. (a) Spectral variation of the mean of the ensemble-averaged Poynting vector norm $\left|\left\langle {\bf S}\left(\mathbf{r}\right)\right\rangle \right|$, (i.e. the light intensity), in a single cylinder  illuminated by a totally coherent GSMS beam. The two magnetic multipole peaks are swhown.  (b) The same quantity in a range of higher $\lambda$ in which the Mie coefficients contributing to the scattering cross section are $b_0$, (electric dipole, $\lambda=6-7 nm$), and $b_1$, (magnetic dipole, $\lambda=2.7 nm$);  hence the particle being magneto-dielectric. The insets in (a) and (b) show the  spatial distribution of $\left|\left\langle {\bf S}\left(\mathbf{r}\right)\right\rangle \right|$ for WGMs:  $TE_{31}$/$WGE_{21}$ and $TE_{11}$/$TE_{01}$, respectively.}
\label{fig: intensity 1particle}
\end{figure}

Hence, the chosen wavelength is the near infrared, (NIR),  in which two multipolar peaks of  field intensity localized  in the cylinder, associated to  its morphology dependent resonances  (MDR) are found, [see Fig. \ref{fig: intensity 1particle}(a)]. As the insets show,  these are  the WGMs $TE_{31}$ ($\lambda\approx 1205nm$) and a $TE_{21}$ ($\lambda\approx 1620nm$). At larger  $\lambda$, as shown  in Fig. \ref{fig: intensity 1particle}(b) the MDRs  $TE_{11}$ ($\lambda\approx 2610nm$) and $TE_{01}$ ($\lambda\approx 6710nm$) are excited, (cf. the insets of this figure). The $TE_{11}$ is interesting because, as shown in  \cite{Peng2007_2,Vynck2009,ValdiviaValero2012_1}, the cylinder scattering cross section is dominated by the Mie coefficients $b_0$ and $b_1$ \cite{vandeHulst1981}, associated to the electric and magnetic dipolar moments, ${\bf p}$ and ${\bf m}$, respectively, of the cylinder; therefore this particle  behaves as  magnetodielectric in this spectral range.

\begin{figure}[htbp]
\begin{minipage}{\columnwidth}
\centering
\includegraphics[width=0.8\linewidth]{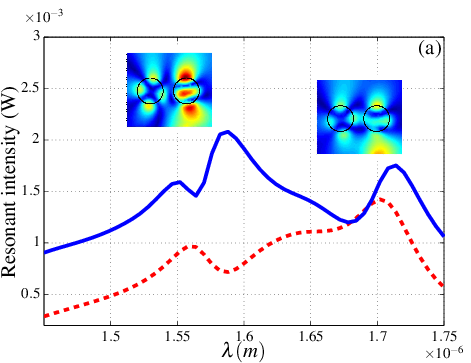}
\end{minipage}
\begin{minipage}{\columnwidth}
\centering
\includegraphics[width=0.8\linewidth]{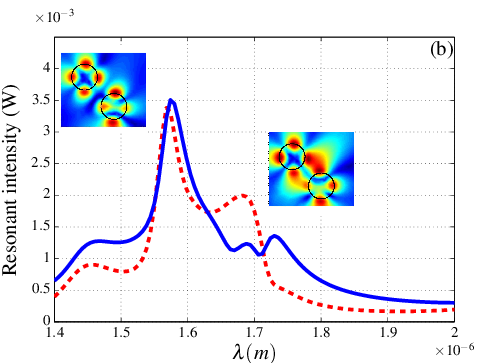}
\end{minipage}
\caption{Spatially coherent illumination. (a)  $\left|\left\langle {\bf S}\left(\mathbf{r}\right)\right\rangle \right|$ localized in each particle of a {\lq\lq}bi-atomic" photonic molecule vs. $\lambda$, illuminated as in Fig. \ref{fig: intensity 1particle}(a). This leads to the splitting of the $TE_{21}$ mode of a single particle, which produces a blue-shifted, (anti-symmetric), and a red-shifted, (symmetric), molecular state, respectively. (b) The same quantity showing the other possibility of splitting associated to the same MDR. The blue solid and red dashed lines in (a) stand for the right, (i.e. the one directly illuminated), and the left particle, respectively. The same code is used in (b), now for the lower, (directly illuminated), and the upper particle, respectively. The insets show the intensity maps of the {\lq\lq}molecular" states, again related  to each intensity peak concentrated by both particles. }
\label{fig: intensity 2particles_4lobes}
\end{figure}

The concentration of intensity $\left|\left\langle {\bf S}\left(\mathbf{r}\right)\right\rangle \right|$ inside each particle conforming the photonic molecule is shown in Figs. \ref{fig: intensity 2particles_4lobes}(a) and  \ref{fig: intensity 2particles_4lobes}(b). A comparison between the blue solid and the red dashed lines in Fig. \ref{fig: intensity 2particles_4lobes}(a) shows that the intensity $\left|\left\langle {\bf S}\left(\mathbf{r}\right)\right\rangle \right|$ in the right particle is generally higher than that in the left one, [the same happens for the lower and the upper cylinders in \ref{fig: intensity 2particles_4lobes}(b)]. This happens because the particle directly illuminated by the beam concentrates more intensity $\left|\left\langle {\bf S}\left(\mathbf{r}\right)\right\rangle \right|$. 

The calculation is focused on the different non-degenerate collective states that can produce the mode $TE_{21}$ excited in both cylinders. Due to the disposition of the lobes of the resonance, ({\lq\lq}even", $E$, or {\lq\lq}odd", $O$ in the field {\bf E} spatial distribution), for each particle with respect to the symmetry axes defined by the ensemble, which are longitudinal and transversal with respect to the molecule axis (hereafter denoted as $Y_{pm}$ and $X_{pm}$, respectively), such a resonance excited in this configuration can generate four {\lq\lq}molecular" states \cite{Boriskina2007,Boriskina2006,Boriskina2010,Boriskina2006_3}. The collective states $Y_{pm}/E$ can be obtained by illuminating the ensemble either in the direction parallel or transversal to the molecule axis. 

The reason to select the configuration in which the $Y_{pm}$ axis appears inclined by an angle $\pi/2$ while the direction of the beam is parallel to the $Y$ axis of the calculation window,  is explained in Subsection \ref{subsec: Force_2particles}.  Figure \ref{fig: intensity 2particles_4lobes}(a) shows this geometrical configuration, which renders the molecular states $Y_{pm}/E$ as consequence of the splitting of the resonance $TE_{21}$ of the single particle into two new MDRs, associated to the disposition of the lobes with respect to the $X_{pm}$ axis, i.e. $X_{pm}Y_{pm}/OE$ and $X_{pm}Y_{pm}/EE$, at $\lambda=1597nm$ and $\lambda=1665nm$, respectively (see the insets) \cite{Boriskina2007}. 

On the other hand, in order to reproduce the collective states $Y_{pm}/O$, the $Y_{pm}$ axis must be inclined by an angle of $\pi/4$ with respect to the propagation direction of the beam because of the number of intensity lobes for the resonance $TE_{21}$ in the single particle. This is seen in Fig. \ref{fig: intensity 2particles_4lobes}(b), where the molecular states $Y_{pm}/O$ arise as a new splitting of the resonance $TE_{21}$ of the single particle, i.e. $X_{pm}Y_{pm}/OO$ and $X_{pm}Y_{pm}/EO$, at $\lambda=1582nm$ and $\lambda=1693nm$, respectively, (see the detail in this figure).

All the non-degenerate states of this photonic molecule associated to the MDR  $TE_{21}$ in each particle are shown by these two orientations of the ensemble. Both orientations present two collective resonances, the $X_{pm}/O$ and $X_{pm}/E$ being blue- and red- shifted, (i.e. more and less energetic, respectively). This can be explained by the insets of  this figure: the $X_{pm}/O$ states concentrate relatively much more light intensity inside the cylinders than the $X_{pm}/E$ ones. Each set of orientation also reminds the formation either of a simple, [Fig. \ref{fig: intensity 2particles_4lobes}(a)], or a double, [Fig. \ref{fig: intensity 2particles_4lobes}(b)], bond between the particles \cite{Boriskina2007}.

\begin{figure}[htbp]
\begin{minipage}{\columnwidth}
\centering
\includegraphics[width=0.8\linewidth]{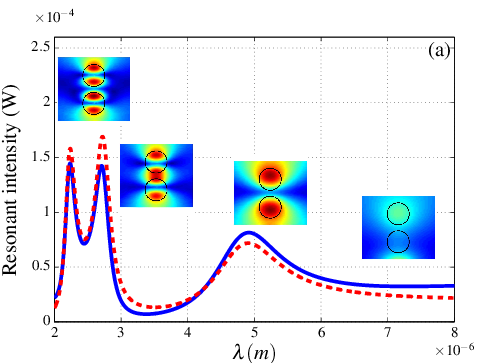}
\end{minipage}
\begin{minipage}{\columnwidth}
\centering
\includegraphics[width=0.8\linewidth]{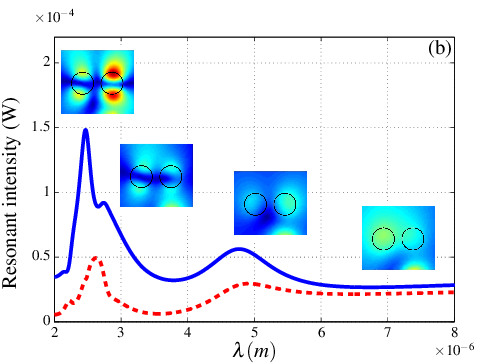}
\end{minipage}
\caption{(a) The same as in Fig. \ref{fig: intensity 2particles_4lobes}(a) in the spectral range in which the single particle is magneto-dielectric, [cf. Fig. \ref{fig: intensity 1particle}(b)]. The first two peaks from the left are associated to the WGM: $TE_{11}$, while the third one is related to the $TE_{01}$ mode. (b) The same as in Fig. \ref{fig: intensity 2particles_2and1lobes}(a) showing the other possibility of splitting for the same MDRs. The interpretation of the so formed {\lq\lq}molecular" states is similar to that of Fig. \ref{fig: intensity 2particles_4lobes}(a) and Fig. \ref{fig: intensity 2particles_4lobes}(b).}
\label{fig: intensity 2particles_2and1lobes}
\end{figure}

By increasing the wavelength $\lambda$ of illumination on this particle pair around the same range as in Fig. \ref{fig: intensity 1particle}(b), the behavior of the collective resonances appears to be similar to that of Fig. \ref{fig: intensity 2particles_4lobes}(a) and Fig. \ref{fig: intensity 2particles_4lobes}(b) regarding the connection between their symmetry, ($X_{pm}/O$ and $X_{pm}/E$ lobes in {\bf E}), and energy, (blue- and red-shifted peaks). These states being in this case originated by the $TE_{11}$ and $TE_{01}$ resonances excited in the single particle. Aiming to reproduce its $Y_{pm}/E$ and $Y_{pm}/O$ states, the $Y_{pm}$ axis is constrained to be either parallel, [see Fig. \ref{fig: intensity 2particles_2and1lobes}(a)], or  perpendicular, [see Fig. \ref{fig: intensity 2particles_2and1lobes}(b)], to the direction of the light beam, respectively. The suppression, in both orientations,  of the less energetic molecular state associated to the $TE_{01}$ WGM of the single particle, i.e.the $X_{pm}Y_{pm}/EE$, is due to the fact that the illuminating  wavelength $\lambda$  is much longer than the dimensions of the molecule, this latter now being almost invisible.

\subsection{Effects of partially coherent illumination on the electromagnetic forces between the particles. Bonding and anti-bonding {\lq\lq}molecular" states}
\label{subsec: Force_2particles}

Next, we consider the cylinder pair illuminated by a GSMS  with different coherence lengths $\sigma_g$. This allows us to understand its effect of the electromagnetic forces acting on its collective states. As previously remarked, for this S-polarization  the fields associated to these states, although localized inside the particles, reach high intensity  values in the area immediately outside them. Taking into account the calculation from the CMR of MST, Eq. (\ref{F_Tij}), maximum forces are thus expected to appear when these states are excited. Two of the MDRs of the single particle: $TE_{21}$ and $TE_{11}$, are selected to study the electromagnetic forces acting in the optical binding between the two  cylinders which conform the photonic molecule.  As discussed in Section \ref{subsec: Intensity_2particles}, each of these resonances splits into two collective states whose symmetry and energy are related to each other. The $TE_{21}$ mode is chosen due to its possibility to generate states in the particle pair which remind those of a simple [Fig. \ref{fig: intensity 2particles_4lobes}(a)] and a double [Fig. \ref{fig: intensity 2particles_4lobes}(b)] bond in an atomic molecule. The $TE_{11}$  mode causes the particles to behave as magneto-dielectric, giving rise to an interaction not only between its electric dipoles,  but also between its induced magnetic ones.

\begin{figure}[htbp]
\begin{minipage}{\columnwidth}
\centering
\includegraphics[width=\linewidth]{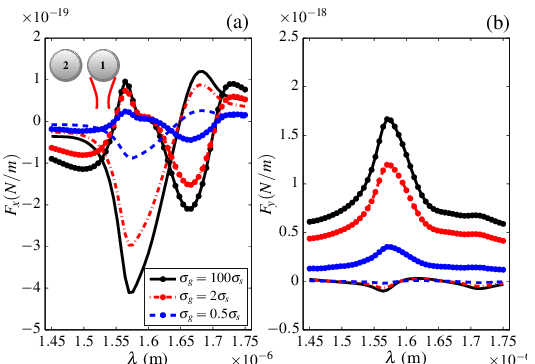}
\end{minipage}
\begin{minipage}{\columnwidth}
\centering
\includegraphics[width=\linewidth]{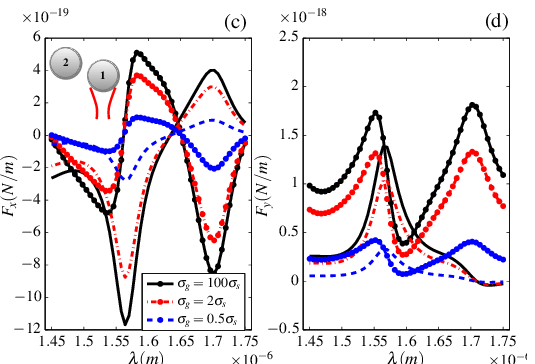}
\end{minipage}
\caption{(a) Horizontal and (b)vertical components of the time-averaged electromagnetic forces per axial  unit length on each cylinder of the particle pair for the orientation shown in Fig. \ref{fig: intensity 2particles_4lobes}(a).  (c)-(d) The same quantities for the molecule oriented according to Fig. \ref{fig: intensity 2particles_4lobes}(b). The lines with and without points  correspond to the force on the particle 1 and 2, respectively. The colors are associated to an illuminating GSMS beam with different  coherence length-to-spot size ratios: $\sigma_g/\sigma_s$: $\sigma_g=100\sigma_s$ (black), $\sigma_g=2\sigma_s$ (red), and $\sigma_g=0.5\sigma_s$ (blue).}
\label{fig: forces_2particles_4lobes}
\end{figure}

Figures \ref{fig: forces_2particles_4lobes}(a)-(b) and \ref{fig: forces_2particles_4lobes}(c)-(d) show the electromagnetic force between the two particles in the case of the collective states corresponding to the two first peaks of intensity $\left|\left\langle {\bf S}\left(\mathbf{r}\right)\right\rangle \right|$ in Fig. \ref{fig: intensity 2particles_4lobes}(a) and Fig. \ref{fig: intensity 2particles_4lobes}(b), respectively. They correspond to the splitting of the magnetic quadrupole $b_2$ of the single particle of Fig. 4(a). The reason to choose the orientation shown in Fig. \ref{fig: forces_2particles_4lobes}(a) for the molecule is now clear since the total force on the particles has two contributions: the gradient force between the particles and that of scattering related to the radiation pressure of the incident beam along $OX$ and $OY$.  On the other hand, the orientation used in Figs. \ref{fig: forces_2particles_4lobes}(c)-(d) causes those two force components to mix with each other  along $OY$, notwithstanding remaining possible   to study the interaction between both particles  by means of  the force X-component.

Under completely coherent illumination, peaks of repulsive and attractive force between the two  particles appear at $\lambda\approx 1597nm$ and $\lambda\approx 1665nm$, (cf.  in Fig. \ref{fig: forces_2particles_4lobes}(a) black lines with and without points  for the X-component of the forces on particles 1 and 2, respectively]. The same happens in Fig. \ref{fig: forces_2particles_4lobes}(c) at $\lambda\approx 1582nm$ and $\lambda\approx 1693nm$. These results allow to identify the blue-shifted $X_{pm}Y_{pm}/OE$ and the red-shifted $X_{pm}Y_{pm}/EE$ in Fig. \ref{fig: intensity 2particles_4lobes}(a) collective states ($X_{pm}Y_{pm}/OO$ and $X_{pm}Y_{pm}/EO$ in Fig. \ref{fig: intensity 2particles_4lobes}(b)) as anti-bonding and bonding ones, respectively \cite{Boriskina2007,Boriskina2006,Boriskina2010,Boriskina2006_3,nieto2004near}.

 The forces in the vertical direction are higher for particle 1 (which is directly illuminated) in both orientations. In Fig. \ref{fig: forces_2particles_4lobes}(a) this component, associated to the scattering force from the beam, is lower for the bonding molecular state at $\lambda\approx 1665nm$ than that for the antibonding one (at $\lambda\approx 1597nm$), since the former  renders higher values of field intensity  immediately outside the particles. For the orientation of Fig. \ref{fig: forces_2particles_4lobes}(c), both collective states, the repulsive and the attractive one,  at $\lambda\approx 1582nm$ and $\lambda\approx 1693nm$, suffer comparable Y-components of the total force because now in this direction the gradient force between the particles must also be taken into account.  

When we decrease the coherence length of the source, (see red and blue lines standing for $\sigma_g=2\sigma_s$ and $\sigma_g=0.5\sigma_s$, respectively), both components of  the force invariably diminish. Although the dimension of the molecule and its position with respect to the source, which is in the lower boundary of the calculation window), are subwavelength, these results are opposite to those of Fig. 4 in \cite{aunon2012photonic}, the interaction between the GSMS beam and the particles now being more complex due to the addition of the effect from the MDRs. In fact, the intensity pattern of the interference process which renders the particle resonance decreases, i.e. the field lobes corresponding to the formation of the resonance in each particle loose contrast. This leads, taking into account the force calculation, to a decrement in the field intensity values reached outside the particles and hence in their  optical attraction or repulsion. 

\begin{figure}[htbp]
\begin{minipage}{\columnwidth}
\centering
\includegraphics[width=\linewidth]{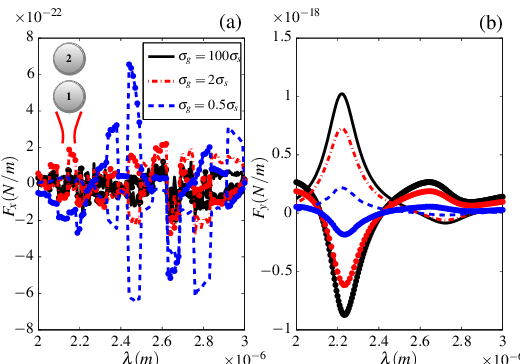}
\end{minipage}
\begin{minipage}{\columnwidth}
\centering
\includegraphics[width=\linewidth]{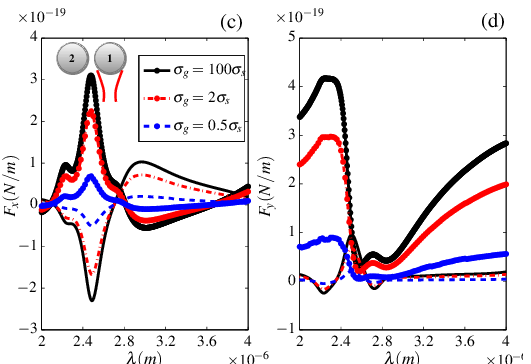}
\end{minipage}
\caption{(a)-(b) The same quantities as in Fig. \ref{fig: forces_2particles_4lobes}(a)-(b) with the molecule oriented as in Fig. \ref{fig: intensity 2particles_2and1lobes}(a). (c)-(d) The same as in Fig. \ref{fig: forces_2particles_2lobes}(a)-(b), the molecule now being oriented as in Fig. \ref{fig: intensity 2particles_2and1lobes}(b). The code of lines and colors is identical to that of Fig. \ref{fig: forces_2particles_4lobes}}
\label{fig: forces_2particles_2lobes}
\end{figure}

Finally, the optical forces on the molecular states associated to the first two peaks of Fig. \ref{fig: intensity 2particles_2and1lobes}(a) and Fig. \ref{fig: intensity 2particles_2and1lobes}(b), [associated to the magnetic dipole of Fig. 4(b)],  are  shown in Figs. \ref{fig: forces_2particles_2lobes}(a)-(b) and \ref{fig: forces_2particles_2lobes}(c)-(d), respectively. The vertical orientation of the pair, although now mixing both contributions to the total force, (i.e. gradient component  between the particles and scattering one due to radiation pressure of the light beam), renders its Y-component being the only  significant  one, and behaves just as expected at $\lambda\approx 2230nm$ and $\lambda\approx 2680nm$. Namely, repulsive and attractive forces arise acting on the blue-shifted, $X_{pm}Y_{pm}/OE$, and  the red-shifted, $X_{pm}Y_{pm}/EE$, collective states, respectively [see Fig. \ref{fig: forces_2particles_2lobes}(a)]. The  force X-component remains null because of  the orientation of the pair.  On the other hand, the horizontal orientation, [see Fig. \ref{fig: forces_2particles_2lobes}(c)], deserves the same discussion on both componets of the total force as that  concerning  Fig. \ref{fig: forces_2particles_4lobes}(a)-(b): repulsive and attractive forces between the particles  now emerge at $\lambda\approx 2480nm$ and $\lambda\approx 2985nm$, which  correspond to the blue-shifted, $X_{pm}Y_{pm}/OO$, and the red-shifted, $X_{pm}Y_{pm}/EO$, molecular states. As in the previous case, the loss of coherence in the light beam causes the decrement in the magnitude of both force components.

\section{Conclusions}
\label{conclusion}

We have presented a theory, illustrated by computer simulations, of optical binding of Mie dipolar  dielectric particles with magnetodielectric behavior,  in the near field of a partially coherent  Gaussian-Schell model source. In connection with previous work \cite{Lindberg2006}, a straightforward representation   has been chosen in the framework of the  angular spectrum \cite{nietolibro,Wolf1982newtheory} and the  coherent mode decomposition \cite{mandel1995optical}. The excitation of the electric dipole and the magnetic dipole and multipoles confers to these systems a rich landscape of resonant forces. For adjusted parameters of the emitted wavefield, i.e. the spot size $\sigma_s$ and the coherence length $\sigma_g$, in contrast with to far-field effects, (see e.g. \cite{wang2007effects, aunon2012opticalforces}),  and confirming other near field results \cite{aunon2012photonic}, as  the  coherence length  $\sigma_g$ decreases,  the pulling force from the source on a single particle increases.

In addition, on extending the analysis to the dynamical interactions between the emitted  light  and a pair of cylinders forming a photonic molecule, we have shown the effects of the spatial coherence on the optical binding beween the particles. This is linked to the symmetric and anti-symmetric molecular resonances, associated to bonding and anti-bonding states, respectively. The role of the interplay between the electric and magnetic induced dipoles when such Mie resonances are induced, has been shown to be important.  Now the threshold of evanescent wave contribution to scattered field is critical. Namely, in addition to being at subwavelength distances from the source plane, the particles need to be practically in contact with each other for a sustantial contribution of the inhomogeneous modes. As a consequence, as few evanescent modes are present, a decrease of the coherence length $\sigma_g$ conveys lower  bonding and antibonding forces.

All this confims that the near field force strength is linked to evanescent waves and increases with a decrease of the near field coherence length, contrary to the effect in the far field where only propagating modes are present.
\section*{Appendix}
\label{Appendix}

We illustrate  force calculations based on the coherent mode representation (CMR)  of Section \ref{one_dimansional_GSMS}.   We  address a cylinder of radius $\lambda/100$, made of Silica glass ($\varepsilon_p=2.1$), illuminated by a  GSMS placed at $y=0$, (cf. Eq. (\ref{Wzz_onedim})). The distance between the source and the center of the particle is $\lambda/10$. The number of modes is determined by the ratio $\sigma_g/\sigma_s$,  the first mode ($q=0$) corresponding to  the globally spatial coherent case studied in Section \ref{Near_Field_forces}). The scheme of the simulation window in which the beam propagates and is scattered by the particle, as well as the method to calculate the optical forces, is similar to that previously explained in Section \ref{fig: Numerical_setup}, now for a single particle.

\begin{figure}[h!]
\begin{centering}
\includegraphics[width=\linewidth]{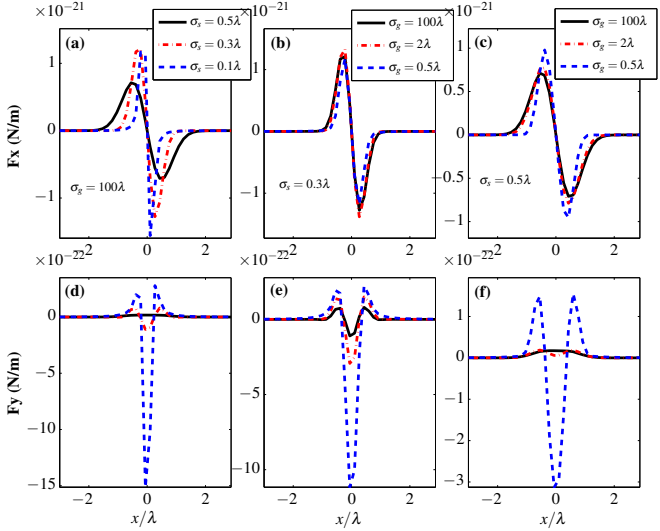}
\par\end{centering}
\caption{(Color online). Ensemble-averaged forces  $F_x$, (first row), and $F_y$, (second row), from a partially coherent GSMS. The first column from the left pertains to the fully coherent source, ($\sigma_g=100\lambda \gg \sigma_s$), which would correspond to the case of Section \ref{OF_GSMS}. For the center and right columns $\sigma_s=0.3\lambda$ and $0.5\lambda$,   respectively}.
\label{fig_fuerza_CMR}
\end{figure}

Fig. \ref{fig_fuerza_CMR} displays the calculated force components . Here  one cannot separate the conservative and non-conservative components of the force since  in Eq. (\ref{F_Tij}) the MST flow yields the total force. Each row of Fig. \ref{fig_fuerza_CMR} represents the ensemble-averaged  forces $\left\langle F_x \right\rangle $ and $\left\langle F_y\right\rangle $ for different values of $\sigma_s$ and $\sigma_g$, (see the legend of the figure). The first column, [Figs. \ref{fig_fuerza_CMR}(a), (d)],  contains $\left\langle F_x\right\rangle $ and $\left\langle F_y\right\rangle $ for the same parameters of  Figs. \ref{fig_gradient} and  \ref{fig_scatt}, (by inverting the color - line code). We see that for a fully coherent source, as we decrease the value of the spot size $\sigma_s$,  the magnitude of the force increases, as stated in the main text. We also observe how $\left\langle F_y\right\rangle $  is negative, (i.e. the particle is pulled to the plane of the source), for $\sigma_s<0.3\lambda$; this is due to the contribution of the evanescent waves. In the main text this fact is discussed.

The second column, [(Figs. \ref{fig_fuerza_CMR} (b), (e)],  represents the force due to a partially coherent GSMS. We have fixed the spot size to $\sigma_s=0.3\lambda$. Contrary to what one could expect, it is the most incoherent emitted field that which produces the maximum force.  In the last column, although we can see a similar behavior, we also observe that for $\sigma_g>0.5\lambda$ the force is positive, i.e., the particle is pushed by the source towards $y>0$.

\begin{figure}[h!]
\begin{centering}
\includegraphics[width=\linewidth]{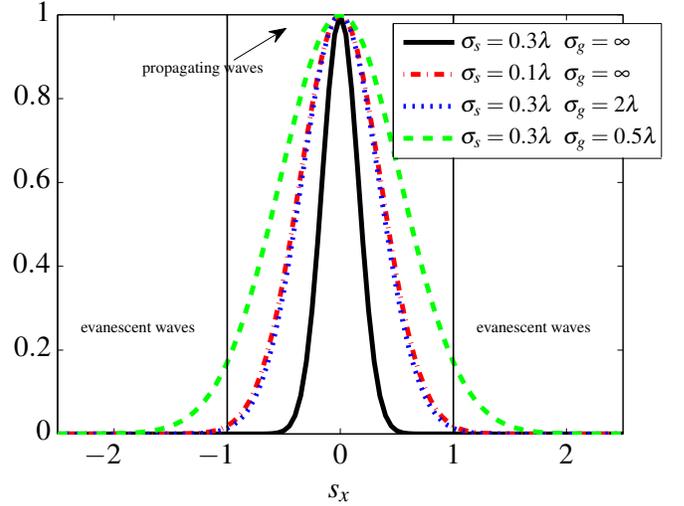}
\par\end{centering}
\caption{(Color online).The function  $\text{exp}(-k^{2}s_{x}^{2}/(4c^2))$ versus the transversal component $s_x$ for different values of the spot size $\sigma_g$ and  coherence length $\sigma_s$. For $s_x>1$ the evanescent waves are not negligible.}
\label{fig_espectro}
\end{figure}

In order to explain all of these results, in Fig. \ref{fig_espectro} we show the exponential function $\text{exp}[-k^{2}s_{x}^{2}/(4c^2)]$  of the angular spectrum, [cf. Eq. (\ref{eigenfunction_angular})], for different values of $\sigma_g$ and $\sigma_s$; this helps us to understand the behavior of the previous figures. The black and the blue point lines represent the width of the Gaussian function for two cases represented in the force in Figs. \ref{fig_fuerza_CMR}(a) and (d). One sees that for a fully coherent source, the Gaussian is broader for a lower value of the spot size, thus taking more evanescent modes of the angular spectrum. The red point-dashed line and the green-dashed lines represent two cases of  Figs. \ref{fig_fuerza_CMR}(b) and  (e). Now, for a partially coherent source, fixing the value of the spot size,  the evanescent modes are more important as  the coherence length of the source decreases. All this agrees with the results of Fig. \ref{fig_fuerza_CMR} and Fig. 4 of \cite{aunon2012photonic}.

\section*{Acknowledgments}
Work supported by the Spanish  Ministerio de Economia y Competitividad (MINECO) through FIS2012-36113-C03-03  research grant.  
JMA  and FJV-V thank a  MINECO  and a Consolider-Nanolight fellowship, respectively. 


\begin{thebibliography}{10}
\newcommand{\enquote}[1]{``#1''}

\bibitem{wang2007effects}
L.~G. Wang, C.~L. Zhao, L.~Q. Wang, X.~H. Lu, and S.~Y. Zhu, \enquote{{Effect
  of spatial coherence on radiation forces acting on a Rayleigh dielectric
  sphere},} Opt. Lett. \textbf{{\bf 32}}, 1393--1395 (2007).

\bibitem{zha2009radiation}
C.~Zhao, Y.~Cai, X.~Lu, and H.~T. Eyyubo\u{g}l, \enquote{{Radiation force of
  coherent and partially coherent flat-topped beams on a Rayleigh particle},}
  Opt. Express \textbf{{\bf 17}}, 1753--1765 (2009).

\bibitem{korotkova2009twisted}
C.~Zhao, Y.~Cai, and O.~Korotkova, \enquote{{Radiation force of scalar and
  electromagnetic twisted Gaussian Schell-model beams},} Opt. Express
  \textbf{17}, 21472--21487 (2009).

\bibitem{gbur}
S. M. Kim and  G. Gbur,  \enquote{{Momentum conservation in partially coherent waveﬁelds},}  Phys.Rev.  A {\bf 79}, 033844 (2009).

\bibitem{aunon2012opticalforces}
J.~M. Au{\~n}{\'o}n and M.~Nieto-Vesperinas, \enquote{{Optical forces on small
  particles from partially coherent light},} J. Opt. Soc. Am. A \textbf{29},
  1389--1398 (2012).

\bibitem {Greffet} C. Henkel, J. Joulain, J.P. Mulet and J. J. Greffet, \enquote{{Radiation forces on small particles in thermal near fields},}  J. Opt. A {\bf 4}, s109-s114 (2002).

\bibitem {Antezza} M. Antezza, L. Pitaevskii and S. Stringari, \enquote{{New asymptotic beahvior  of the surface-atom force out of thermal equilibrium},} Phys. Rev. Lett. {\bf 95}, 113202 (2005).

\bibitem{aunon2012photonic}
J.~M. Au{\~n}{\'o}n and M.~Nieto-Vesperinas, \enquote{{Photonic forces in the
  near field of statistically homogeneous fluctuating sources},} Phys. Rev. A
  \textbf{85}, 053828 (2012).

\bibitem{aunon2013photonic2} J.~M. Au{\~n}{\'o}n, C.W. Qiu  and M.~Nieto-Vesperinas, \enquote{{Tailoring photonic forces on a magnetodielectric nanoparticle with a fluctuating optical source},} Phys. Rev. A
  \textbf{88}, 043817 (2013).

\bibitem{Dogariu} S. Sukhov,  K. Douglass and A.  Dogariu, "Dipole - dipole interaction in random electromagnetic fields", Opt. lett. {\bf  38}, 2385 (2013).

\bibitem{Chaumet1} P. Chaumet and M. Nieto-vesperinas, "Optical binding of particles with or without the presence of a ﬂat dielectric surface", Phys. Rev. B {\bf  64}, 035422 (2001).

\bibitem{block2004review}
K.~C. Neuman and S.~M. Block, \enquote{Optical trapping,} Review of Scientific
  Instruments \textbf{75}, 2787--2809 (2004).

\bibitem{povinelli}
M. L. Povinelli, S. G. Johnson, M. Loncar, M. Ibanescu, E. J. Smythe, F. Capasso, and J. D. Joannopoulos, " High-Q enhancement of attractive and repulsive optical forces between coupled whispering-gallery-mode resonators", Opt. Express {\bf 13}, 8286-8295 (2005).

\bibitem{zemanek} K. Dholakia and P. Zemanek, "Gripped by light: Optical binding", Rev. Mod. Opt. {\bf  82}, 1767-1791 (2010).


\bibitem{mandel1995optical}
L.~Mandel and E.~Wolf, \emph{{Optical Coherence and Quantum Optics}} (Cambridge
  U. Press, Cambridge, UK, 1995).

\bibitem{james} D. F. V. James and E. Wolf, “Correlation-induced spectral
changes,” Rep. Prog. Phys. {\bf 59},  771–818 (1996).

\bibitem{Carney1}
P.~S. Carney, E.~Wolf, and G.~S. Agarwal, \enquote{{Statistical generalizations
  of the optical cross-section theorem with application to inverse
  scattering},} J. Opt. Soc. Am. A \textbf{{\bf 14}}, 3366--3371 (1997).

\bibitem{Lindberg2006}
J.~Lindberg, T.~Set\"{a}l\"{a}, M.~Kaivola, and A.~T. Friberg, \enquote{Spatial
  coherence effects in light scattering from metallic nanocylinders,} J. Opt.
  Soc. Am. A \textbf{23}, 1349--1358 (2006).

\bibitem{wolf2007introduction}
E.~Wolf, \emph{{Introduction to the Theory of Coherence and Polarization of
  Light}} (Cambridge U. Press, New York, 2007).

\bibitem{carminati2010subwavelength}
R.~Carminati, \enquote{Subwavelength spatial correlations in near-field speckle
  patterns,} Phys. Rev. A \textbf{81}, 053804 (2010).

\bibitem{Ashkin1970}
A.~Ashkin, \enquote{{Acceleration and trapping of particles by radiation
  pressure},} Phys. Rev. Lett. \textbf{{\bf 24}}, 156--159 (1970).

\bibitem{Gori2002synthesis}
G.~Piquero, F.~Gori, P.~Romanini, M.~Santarsiero, R.~Borghi, and A.~Mondello,
  \enquote{Synthesis of partially polarized gaussian schell-model sources,}
  Optics Communications \textbf{208}, 9 -- 16 (2002).

\bibitem{Friberg1988Imaging}
A.~T. Friberg and J.~Turunen, \enquote{Imaging of gaussian schell-model
  sources,} J. Opt. Soc. Am. A \textbf{5}, 713--720 (1988).

\bibitem{aunon13equivalence}
J.~M. Au{\~n}{\'o}n and M.~Nieto-Vesperinas, \enquote{Partially coherent
  fluctuating sources that produce the same optical force as a laser beam,}
  Opt. Lett. \textbf{38}, 2869--2872 (2013).

\bibitem{Wolf1982Coherentmode}
A.~Starikov and E.~Wolf, \enquote{Coherent-mode representation of gaussian
  schell-model sources and of their radiation fields,} J. Opt. Soc. Am.
  \textbf{72}, 923--928 (1982).

\bibitem{Wolf1982newtheory}
E.~Wolf, \enquote{New theory of partial coherence in the space-frequency
  domain. part i: spectra and cross spectra of steady-state sources,} J. Opt.
  Soc. Am. \textbf{72}, 343--351 (1982).


\bibitem{Peng2007_2}
L.~Peng, L.~Ran, H.~Chen, H.~Zhang, J.~A. Kong, and T.~M. Grzegorczyk,
  \enquote{Experimental observation of left-handed behavior in an array of
  standard dielectric resonators,} Phys. Rev. Lett. \textbf{98}, 157403 (2007).


\bibitem{Vynck2009}
K.~Vynck, D.~Felbacq, E.~Centeno, A.~I. C\ifmmode~\u{a}\else \u{a}\fi{}buz,
  D.~Cassagne, and B.~Guizal, \enquote{All-dielectric rod-type metamaterials at
  optical frequencies,} Phys. Rev. Lett. \textbf{102}, 133901 (2009).

\bibitem{GarciaEtxarri11Anistropic}
A.~Garc{\'i}a-Etxarri, R.~G{\'o}mez-Medina, L.~S. Froufe-P{\'e}rez,
  C.~L{\'o}pez, L.~Chantada, F.~Scheffold, J.~Aizpurua, M.~Nieto-Vesperinas,
  and J.~J. S{\'a}enz, \enquote{{Strong magnetic response of submicron Silicon
  particles in the infrared},} Opt. Express \textbf{\bf 19}, 4815--4826 (2011).

\bibitem{magnelight}
A. I. Kuznetsov, A. E. Miroshnichenko, Y. H. Fu, J. Zhang and B. Luk’yanchuk, "Magnetic light",  Sci. Reports {\bf 2}, 492 (2012).



\bibitem{nietoJOSA011}
M.~Nieto-Vesperinas, R.~Gomez-Medina, and J.~J. S{\'a}enz,
  \enquote{{Angle-suppressed scattering and optical forces on submicrometer
  dielectric particles},} J. Opt. Soc. Am. A \textbf{{\bf 28}}, 54--60 (2011).

\bibitem{nieto2004near}
M.~Nieto-Vesperinas, P.~C. Chaumet, and A.~Rahmani, \enquote{{Near-field
  photonic forces},} Phil. Trans. R. Soc. Lond. A \textbf{{\bf 362}}, 719--737
  (2004).

\bibitem{wong2006gradient}
V.~Wong and M.~A. Ratner, \enquote{{Gradient and nongradient contributions to
  plasmon-enhanced optical forces on silver nanoparticles},} Phys. Rev. B
  \textbf{{\bf 73}}, 075416 (2006).

\bibitem{nieto2010optical}
M.~Nieto-Vesperinas, J.~J. S{\'a}enz, R.~G{\'o}mez-Medina, and L.~Chantada,
  \enquote{{Optical forces on small magnetodielectric particles},} Opt. Express
  \textbf{{\bf 18}}, 11428--11443 (2010).

\bibitem{nietolibro}
M.~Nieto-Vesperinas, \emph{{Scattering and Diffraction in Physical Optics}}
  (World Science, Singapur, 2006).

\bibitem{nietoconj} 
M. Nieto-Vesperinas and E. Wolf, “Phase conjugation and symmetries with wave fields in free space containing evanescent components”,  J.  Opt. Soc. Am. A  {\bf 2} ,    1429-1434   (1985).

\bibitem{nietolhm}
M. Nieto-Vesperinas, “Problem of image superresolution with a negative-refractive-index slab“,   J.  Opt. Soc. Am. A  {\bf 21} ,    491-498   (2004).

\bibitem{carminati1999near}
R.~Carminati and J.-J. Greffet, \enquote{Near-field effects in spatial
  coherence of thermal sources,} Phys. Rev. Lett. \textbf{82}, 1660--1663
  (1999).

\bibitem{setala2002near}
T.~Set{\"a}l{\"a}, A.~Shevchenko, M.~Kaivola, and A.~T. Friberg,
  \enquote{{Degree of polarization for optical near fields},} Phys. Rev. E
  \textbf{66}, 016615 (2002).

\bibitem{Ellis2005degree}
J.~Ellis, A.~Dogariu, S.~Ponomarenko, and E.~Wolf, \enquote{{Degree of
  polarization of statistically stationary electromagnetic fields},} Opt.
  Commun. \textbf{248}, 333--337 (2005).

\bibitem{Ellis2005degree2}
J.~Ellis and A.~Dogariu, \enquote{On the degree of polarization of random
  electromagnetic fields,} Optics Communications \textbf{253}, 257 -- 265
  (2005).

\bibitem{aunon2013degree}
J.~M. Au{\~n}{\'o}n and M.~Nieto-Vesperinas, \enquote{On two definitions of the
  three-dimensional degree of polarization in the near field of statistically
  homogeneous partially coherent sources,} Opt. Lett. \textbf{38}, 58--60
  (2013).

\bibitem{tervo2004theory}
J.~Tervo, T.~Set{\"a}l{\"a}, and A.~T. Friberg, \enquote{Theory of partially
  coherent electromagnetic fields in the space--frequency domain,} J. Opt. Soc.
  Am. A \textbf{21}, 2205--2215 (2004).


\bibitem{jackson1998classical}
J.~D. Jackson, \emph{{Classical Electrodynamics}} (Wiley, New York, New York,
  1998).

\bibitem{ChaumetOL}
P.~C. Chaumet and M.~Nieto-Vesperinas, \enquote{{Time-averaged total force on a
  dipolar sphere in an electromagnetic field},} Opt. Lett. \textbf{{\bf 25}},
  1065--1067 (2000).

\bibitem{Cui2008}
X.~Cui, D.~Erni, and C.~Hafner, \enquote{Optical forces on metallic
  nanoparticles induced by a photonic nanojet,} Opt. Express \textbf{16},
  13560--13568 (2008).

\bibitem{Gori1980Colletwolf}
F.~Gori, \enquote{Collett-Wolf sources and multimode lasers,} Optics
  Communications \textbf{34}, 301 -- 305 (1980).

\bibitem{Boriskina2006}
S.~V. Boriskina, \enquote{Theoretical prediction of a dramatic q-factor
  enhancement and degeneracy removal of whispering gallery modes in symmetrical
  photonic molecules,} Opt. Lett. \textbf{31}, 338--340 (2006).

\bibitem{Boriskina2007}
S.~V. Boriskina, T.~M. Benson, and P.~Sewell, \enquote{Photonic molecules made
  of matched and mismatched microcavities: new functionalities of microlasers
  and optoelectronic components,} in \enquote{Proc. SPIE,} , vol. 6452 (2007),
  vol. 6452, p. 64520X.

\bibitem{Boriskina2010}
S.~Boriskina, \enquote{Photonic molecules and spectral engineering,} in
  \enquote{Photonic Microresonator Research and Applications,} , vol. 156 of
  \emph{Springer Series in Optical Sciences}, I.~Chremmos, O.~Schwelb, and
  N.~Uzunoglu, eds. (Springer US, 2010), pp. 393--421.

\bibitem{vandeHulst1981}
 H.~van~de Hulst, \emph{Light scattering by small particles}
  (Dover Pubns, 1981).

\bibitem{Taflove2004}
Z.~Chen, A.~Taflove, and V.~Backman, \enquote{Photonic nanojet enhancement of
  backscattering of light by nanoparticles: a potential novel visible-light
  ultramicroscopy technique,} Opt. Express \textbf{12}, 1214--1220 (2004).

\bibitem{Ho1994}
M.~K. Chin, D.~Y. Chu, and S.-T. Ho, \enquote{Estimation of the spontaneous
  emission factor for microdisk lasers via the approximation of whispering
  gallery modes,} Journal of Applied Physics \textbf{75}, 3302--3307 (1994).

\bibitem{ValdiviaValero2012_1}
F.~Valdivia-Valero and M.~Nieto-Vesperinas, \enquote{Composites of resonant
  dielectric rods: A test of their behavior as metamaterial refractive
  elements,} Photonics and Nanostructures - Fundamentals and Applications
  \textbf{10}, 423 -- 434 (2012).

\bibitem{ValdiviaValero2012_2}
F.~Valdivia-Valero and M.~Nieto-Vesperinas, \enquote{Optical forces on
  cylinders near subwavelength slits: effects of extraordinary transmission and
  excitation of Mie resonances,} Optics Express \textbf{20}, 13368--13389
  (2012).

\bibitem{ValdiviaValero2010}
F.~J. Valdivia-Valero and M.~Nieto-Vesperinas, \enquote{Resonance excitation
  and light concentration in sets of dielectric nanocylinders in front of a
  subwavelength aperture. effects on extraordinary transmission,} Opt. Express
  \textbf{18}, 6740--6754 (2010).


\bibitem{ValdiviaValero2011_3}
F.~J. Valdivia-Valero and M.~Nieto-Vesperinas, \enquote{Propagation of particle
plasmons in sets of metallic nanocylinders at the exit of subwavelength
slits,} Journal of Nanophotonics \textbf{5}, 053520--053520--15 (2011).



\bibitem{Boriskina2006_3}
S.~V. Boriskina, \enquote{Spectrally engineered photonic molecules as optical
  sensors with enhanced sensitivity: a proposal and numerical analysis,} J.
  Opt. Soc. Am. B \textbf{23}, 1565--1573 (2006).

\end{thebibliography}

\end{document}